\documentclass[preprint,aip,jcp]{revtex4-1}

\usepackage{graphicx}
\usepackage[mathscr]{eucal}
\usepackage{amssymb}
\usepackage{amsmath}
\usepackage{hyperref}
\usepackage[utf8]{inputenc} 

\begin{document}

\title{Microcanonical characterization of first-order phase transitions in a generalized model for aggregation}

\author{L. F. Trugilho}%
\author{L. G. Rizzi}%
\affiliation{Departamento\,de\,F\'isica, Universidade\,Federal\,de\,Vi\c{c}osa\,(UFV), \\
Av.\,P.\,H.\,Rolfs,~s/n,~36570-900,~Vi\c{c}osa,~Brazil.}

\begin{abstract}
Aggregation transitions in disordered mesoscopic systems play an important role in several areas of knowledge, from materials science to biology.
	The lack of a thermodynamic limit in systems that are intrinsically finite makes the microcanonical thermostatistics analysis, which is based on the microcanonical entropy, a suitable alternative to study the aggregation phenomena.
	Although microcanonical entropies have been used in the characterization of
first-order phase transitions in many non-additive systems, most of the studies are only done numerically with aid of advanced Monte Carlo simulations.
	Here we consider a semi-analytical approach to characterize aggregation transitions that occur in a generalized model related to the model introduced by Thirring.
	By considering an effective interaction energy between the particles in the aggregate, 
our approach allowed us to obtain scaling relations not only for the microcanonical entropies and temperatures, but also for the sizes of the aggregates and free-energy profiles.
	In addition, we test the approach commonly used in simulations which is based on the conformational microcanonical entropy determined from a density of states that is a function of the potential energy only.
	Besides the evaluation of temperature versus concentration phase diagrams, we explore this generalized model to illustrate how one can use the microcanonical thermostatistics as an analysis method to determine experimentally relevant quantities such as latent heats and free-energy barriers of aggregation transitions.
\end{abstract}

\maketitle

\section{Introduction}
\label{intro}

	Since the early works on atomistic cluster transitions~\cite{labastie1990prl,wales1995jcp,grossbook}
and ensemble inequivalence~\cite{barre2001prl,chomaz2006eurphysjA},
	microcanonical thermostatistics analysis has become an important 
approach in the characterization of phase transitions that occur in intrinsically finite systems~\cite{schnabel2011pre,zierenberg2016polymers,jankepaul,qibachmann2018prl}.
	Numerical simulations based on advanced Monte Carlo algorithms 
like multicanonical~\cite{berg1992prl,berg2003cpc}, 
entropic sampling~\cite{lee1993prl},
broad histogram~\cite{broadhist1996brazjphys},
Wang-Landau~\cite{wanglandau2001prl},
and 
statistical temperature~\cite{straub2011jcp,rizzi2011jcp},
	have been instrumental in the evaluation of microcanonical entropies $S(E)$, density 
of states $\Omega=e^{S(E)/k_B}$, and caloric curves $b(E)=k_B^{-1} dS(E)/dE$.
	Examples of computational studies include not only 
lattice~\cite{janke1998nuclphys,chomaz2000prl,pleimling2001jstatphys,pleimling2005pt,beath2006prb,behringer2006pre,martinmayor2007prl,nogawa2011pre}
and magnetic models~\cite{rizzi2016prl,rizzi2011jcis}, but also more sophisticated biopolymeric systems, in particular, models for
peptide aggregation~\cite{junghans2006prl,chen2008pre,junghans2008jcp,junghans2009epl,junghans2011cpc,koci2011pre,trugilho2020jphysconfser},
protein dimerization~\cite{church2012jcp},
homopolymer collapse~\cite{taylor2009pre,taylor2009jcp,taylor2010physproc}, protein folding~\cite{scheraga1994jphyschem,chen2007pre,rojas2008prl,bereau2010jacs,bereau2011jbiophysj,liu2012jcp,frigori2013jcp,frigori2014pre,alves2015cpc,frigori2017pccp,frigori2021jmolmod},
and polymer adsorption~\cite{chen2009jcp,wangliang2009jcp,moddel2010pccp}.

	Theoretical studies involving the microcanonical characterization of
phase transitions, on the other hand, are scarce, and correspond to a limited
number of mean-field-like models~\cite{campa2009physrep}.
	Even so, an interesting analytically tractable model related to aggregation of 
particles is the Thirring's model~\cite{thirring1970zphys}, which was developed in the 
context of gravitational systems.
	Although this model was introduced in the 70s, it has gained renewed interest due to
its non-additive properties caused by its longe-range interactions~\cite{campa2016jstat,latella2015prl}.
	Another important feature of such model is that it displays an aggregation 
transition that is independent of the shape of the forming aggregate, so that it 
is an appealing candidate to describe (at least qualitatively) the microcanonical 
properties of the aggregation transitions observed, for instance, in disordered systems
under confinement~\cite{zierenberg2014jcp,mueller2015physproc,zierenberg2018jphysconfser}.

	It is worth mentioning that microcanonical thermostatistics analysis was recently explored by the numerical methodology presented in Ref.~\cite{janke2017natcommun}, where the authors tried to establish a 
relationship between the shape-free properties of free-energy profiles and caloric curves obtained from microcanonical entropies to the phase transformation kinetics in finite systems with aggregating polymeric and Lennard-Jones particles.
	Interestingly, a similar relationship was proposed in Ref.~\cite{frigori2013jcp} in the context of protein folding, however it were only in Ref.~\cite{rizzi2020jstat} that the free-energy profiles evaluated from the microcanonical entropy was used to derive a kinetic approach that lead to temperature-dependent expressions for the rate constants.
	Hence, the evaluation of semi-analytical methods to obtain caloric curves and microcanonical entropies is of interest for the kinetic approaches developed in Refs.~\cite{rizzi2020jstat,trugilho2021arxiv}.

	In this work, we consider a theoretical approach based on the microcanonical thermostatistics 
analysis to study the aggregation transition described by a generalized version of the 
Thirring's model.
	We obtain expressions for the density of states $\Omega(E)$, from which a comprehensive thermostatistical characterization based on the microcanonical entropy $S(E)$ can be made.
	This characterization demonstrates that the generalized model, just as the usual model, presents first-order phase transitions which correspond to the aggregation of particles that are found in a finite volume.
	Our results are used to demonstrate that the scaling relations inferred from the usual model~\cite{campa2016jstat} for energies and temperatures as functions of the number of particles can be extended to its generalized version, and we show that they lead to additional relations for quantities like entropy, transitions temperatures, latent heats, and free-energy barriers.
	Also, we test the alternative microcanonical approach based on the conformational density of states $\Omega_p(E_p)$, {\it i.e.}, ignoring the contribution from the kinetic energy, which is the most widespread analysis used in Monte Carlo simulations~\cite{schierz2016pre,janke2017jphysconfser}.

\section{Model description}

	In order to define the aggregation model, we consider a system with fixed total volume $V$ with $N$
particles immersed in an implicit solvent. 
	Besides, a number $n$ of these particles is assumed to be aggregated in an arbitrary volume $V_0$, while the other $n'=N-n$ particles are diluted in the remaining volume $V'=V-V_0$.
	We assume that the particles in the diluted phase do not interact with each other, so that their 
statistics is similar to an ideal gas. 
	On the other hand, the $n$ aggregated particles have an effective interaction energy given by
\begin{equation}
E_p(n)=-\nu g(n),
\label{Ep}
\end{equation}
where $g(n)$ is the effective number of bonds and $\nu>0$ is the magnitude of interaction between these particles. 
	Choosing $g(n)\approx n^2$ leads to the usual Thirring's model~\cite{thirring1970zphys,campa2016jstat}, 
where all $n$ particles interact equally with each other.
	In contrast, $g(n)=n-1$ is equivalent to a linear chain with nearest-neighbour interactions. 
	Since we are interested in molecular systems, we assume that the particles at the surface of the aggregate do
not interact equally with those in the ``bulk'', hence we set\footnote{In fact, the possibility of using different exponents $\alpha$ was already considered in Ref.~\cite{thirring1970zphys}.}
\begin{equation} 
g(n)=n^\alpha-1,
\label{gn_alpha}
\end{equation} 
with $1 \leq \alpha \leq 2$.
	In addition to the potential energy, we also consider that the system has kinetic energy 
$E_k$ distributed between all the $N$ particles, so that the total energy of the particles is $E=E_p+E_k$.

\subsection{Density of states}

	Next, we consider the calculation of the density of states\footnote{Although we termed
$\Omega(N,V,E)$ as the density of states, this quantity is defined as
the total number of microscopic states and is related to the ``Gibbs entropy'' $S_G(E)$, 
see Sec.~\ref{boltzmann_vs_gibbs}.}
$\Omega(N,V,E)$ for a system described by our generalized aggregation model.
	By assuming that we are dealing with $N$ classical particles in three dimensions, 
each microscopic state of the system is characterized by the three components of position 
$\vec{r}_i$ and linear momentum $\vec{p}_i$ of each particle 
with $i=1,...,N$.
	Hence, the density of states is obtained by summing over all microscopic states
\begin{equation}
\Omega(E)=\frac{1}{N!\,h^{3N}}\int \text{d}\textbf{q} \, \text{d}\textbf{p} \, \Theta(E-H(\textbf{q},\textbf{p})),
\label{Omega}
\end{equation}
where we use the simplified notation $\textbf{q}=(\vec{r}_1,...,\vec{r}_N)$, $\textbf{p}=(\vec{p}_1,...,\vec{p}_N)$ and the integration is carried out over the entire phase space, {\it i.e.}, over all coordinates of position and momentum of all 
particles;
	here $h$ is the Planck constant, $H(\textbf{q},\textbf{p})$ is the classical Hamiltonian of the $N$-particle system,
 and $\Theta(E-H(\textbf{q},\textbf{p}))$ is the Heaviside (step) function. 
	The volume dependence on the density of states is implicit in the integration limits of the position coordinates in Eq.~\ref{Omega}, but we omit both $N$ and $V$ dependencies at $\Omega(E)$ because 
the following analysis is only based on the energy dependence.

	Since the system is classical, the position and momentum coordinates are independent.
	Also, because the potential energy does not depend on the momentum coordinates, it is possible to integrate 
these coordinates in Eq.~\ref{Omega} separating the Hamiltonian as $H(\textbf{q},\textbf{p})=E_k(\textbf{p})+E_p(\textbf{q})$, with $E_k=(1/2m)\sum_{i=1}^{N}p_i^2$ and $m$ being the mass of each particle.
	This last assumption is valid for a large class of physical systems, including the one modelled here. 
	Integration of the momentum coordinates gives~\cite{pearson1985prA,schierz2015jcp,calvo2000jcp}
\begin{equation}
\Omega(E)=\left(\frac{2\pi m}{h^2}\right)^{3N/2} \frac{1}{N! \, \Gamma(3N/2+1)}\int\text{d}\textbf{q} \, (E-E_p(\textbf{q}))^{3N/2} \,\Theta (E-E_p(\textbf{q})),
\label{Omega_q}
\end{equation} 
where $\Gamma(3N/2+1)$ is the gamma function. 
By considering that the conformational density of states, {\it i.e.}, the number os states with potential energy $E_p$, is defined as $\Omega_p(E_p)=(V^{N})^{-1} \int\text{d}\textbf{q}\,\delta(E_p-E_p(\textbf{q}))$,
it is possible to rewrite Eq.~\ref{Omega_q} as an integral over all the values of the potential energy instead of integrating the positions $\textbf{q}$ of the particles, that is,
\begin{equation}
\Omega(E)= \left(\frac{2\pi m}{h^2}\right)^{3N/2} \frac{V^N}{N! \, \Gamma(3N/2+1)}\int_{-\infty}^{+\infty}\text{d}E_p \, \Omega_p(E_p) \, (E-E_p)^{3N/2} \, \Theta(E-E_p).
\label{Omega_p}
\end{equation} 
	Since the potential energy $E_p(n)$, Eq.~\ref{Ep}, depends only on the natural number $n$ of particles 
in the aggregate, and this dependence is univocal, the integral in Eq.~\ref{Omega_p} can be 
replaced by the sum,
\begin{equation}
\Omega(E)\propto\sum_{n=n_{\min}}^{N}\Omega_p(n) \, (E-E_p(n))^{3N/2},
\label{Omega_E}
\end{equation}
where the multiplicative terms that do not depend on
energy were omitted.
	In Eq.~\ref{Omega_E} the summation is restricted to the interval [$n_{\min},N$] because, if $E<0$, it is necessary that a minimal number of particles be in the aggregated phase, since the kinetic energy $E_k=E-E_p$ must be 
always positive. 
	On the other hand, if $E>0$, then $n_{\min}=1$, which
ensures that $E>E_p$.

	It now remains to obtain the conformational density of states $\Omega_p(n)$,
{\it i.e.}, the number of ways that is possible to arrange $N$ particles so that $n$ are aggregated 
in the volume $V_0$ and $n'=N-n$ are diluted in volume $V'=V-V_0$. 
	This is done considering the probability of finding one particle in $V_0$, $p=V_0/V$, 
and the probability of finding one particle in the volume $V'$, $q=1-p=V'/V$. 
	With that, the probability of finding $n$ particles in $V_0$ and $n'$ in $V'$ is given 
by $p^nq^{n'}=p^nq^{N-n}$. 
	In addition, it is necessary to consider the possible permutations between the particles 
so that $\Omega_p(n)$ is given as
\begin{equation}
\Omega_p(n)=\left(\frac{N}{n}\right)p^nq^{N-n}=\frac{N!}{n! \, (N-n)!}\left(\frac{V_0}{V}\right)^n\left(\frac{V-V_0}{V}\right)^{N-n}.
\label{densidade_n}
\end{equation}
	Following references~\cite{thirring1970zphys,campa2016jstat}, we
define the reduced ``volume'' $\eta$ as
\begin{equation}
\eta=\ln\left(\frac{V-V_0}{V_0}\right).
\label{eta}
\end{equation}
With this definition it is possible to rewrite Eq.~\ref{densidade_n} as
\begin{equation}
\Omega_p(n)=\frac{N!}{n!(N-n)!}\frac{e^{\eta(N-n)}}{(1+e^\eta)^N},
\label{densidade_eta}
\end{equation}
that is, it is possible to express the conformational density of states in terms of a single volume parameter $\eta$, 
which is a control parameter of the model and, as will be shown, it determines whether the system exhibits first-order phase transitions or not.
	Finally, by inserting expression~\ref{densidade_eta} into Eq.~\ref{Omega_E}, we arrive at the final expression for the density of states of our generalized aggregation model,
\begin{equation}
\Omega(E)\propto\sum_{n=n_{\min}}^{N}\frac{e^{\eta(N-n)}}{n!\,(N-n)!}\,\left(E+\nu g(n)\right)^{3N/2},
\label{densidade_E}
\end{equation}
where the terms that do not depend on energy were omitted again, and we used the definition of the 
potential energy, Eq.~\ref{Ep}, with $g(n)=n^\alpha-1$.

\section{Microcanonical thermostatistics}

\subsection{Microcanonical entropy}
\label{entropia}

	Now that we have derived an expression for the density of states $\Omega(E)$,
the microcanonical entropy $S(E)=k_B\ln \Omega(E)$ can be numerically evaluated by performing the summation over $n$ in Eq.~\ref{densidade_E}.
	Alternatively, following references~\cite{thirring1970zphys,campa2016jstat}, one may define a function
$S(E,n)$ so that
\begin{equation}
\Omega(E)=\sum_{n=n_{\min}}^{N}e^{S(E,n)/k_B},
\label{OmegaEsum}
\end{equation}  
where
\begin{equation}
S(E,n)/k_B=
\ln\left[\frac{e^{\eta(N-n)}}{n!\,(N-n)!} \, (E+\nu g(n))^{3N/2}\right] 
+\mathcal{C}(N,V,\eta),
\label{S_n}
\end{equation}
with the constant $\mathcal{C}(N,V,\eta)$ given by the logarithm of the omitted terms in Eq.~\ref{densidade_E}.
	Thus, by carefully considering a Stirling's approximation that is reliable even for small size aggregates~\cite{chesnut1984amjphys}, {\it i.e.}, $n!\approx\sqrt{1+2\pi n}\left(n/e\right)^n$, Eq.~\ref{S_n} leads to
\begin{eqnarray}
S(E,n)/k_B\approx \frac{3N}{2}\ln(E+\nu g(n))+\eta(N-n)-\frac{1}{2}\ln(1+2\pi n)-n\ln n \nonumber \\
~~ ~~ ~~ ~~ -\frac{1}{2}\ln(1+2\pi(N-n)) -(N-n)\ln(N-n)+N +\mathcal{C}(N,V,\eta)/k_B.
\label{entropia_final}
\end{eqnarray}
	From this last expression it is possible to estimate the density of states given by Eq.~\ref{OmegaEsum} as $\Omega(E) \approx e^{S(E,\bar{n})/k_B}$, where the sum over $n$ is replaced by its largest term.
	Hence, we evaluate the microcanonical entropy through an optimization procedure as 
in Ref.~\cite{campa2016jstat}, that is,
\begin{equation}
S(E) \equiv S(E,\bar{n}) \approx k_B \ln \Omega(E),
\label{micro_entropy_def}
\end{equation}
where $\bar{n}=\bar{n}(E)$ is the value of $n$ that maximizes $S(E,n)$ for a given value of total energy $E$
(with $N$ and $V$ fixed).

\begin{figure*}[!t]
	\centering
	\includegraphics[width=0.95\textwidth]{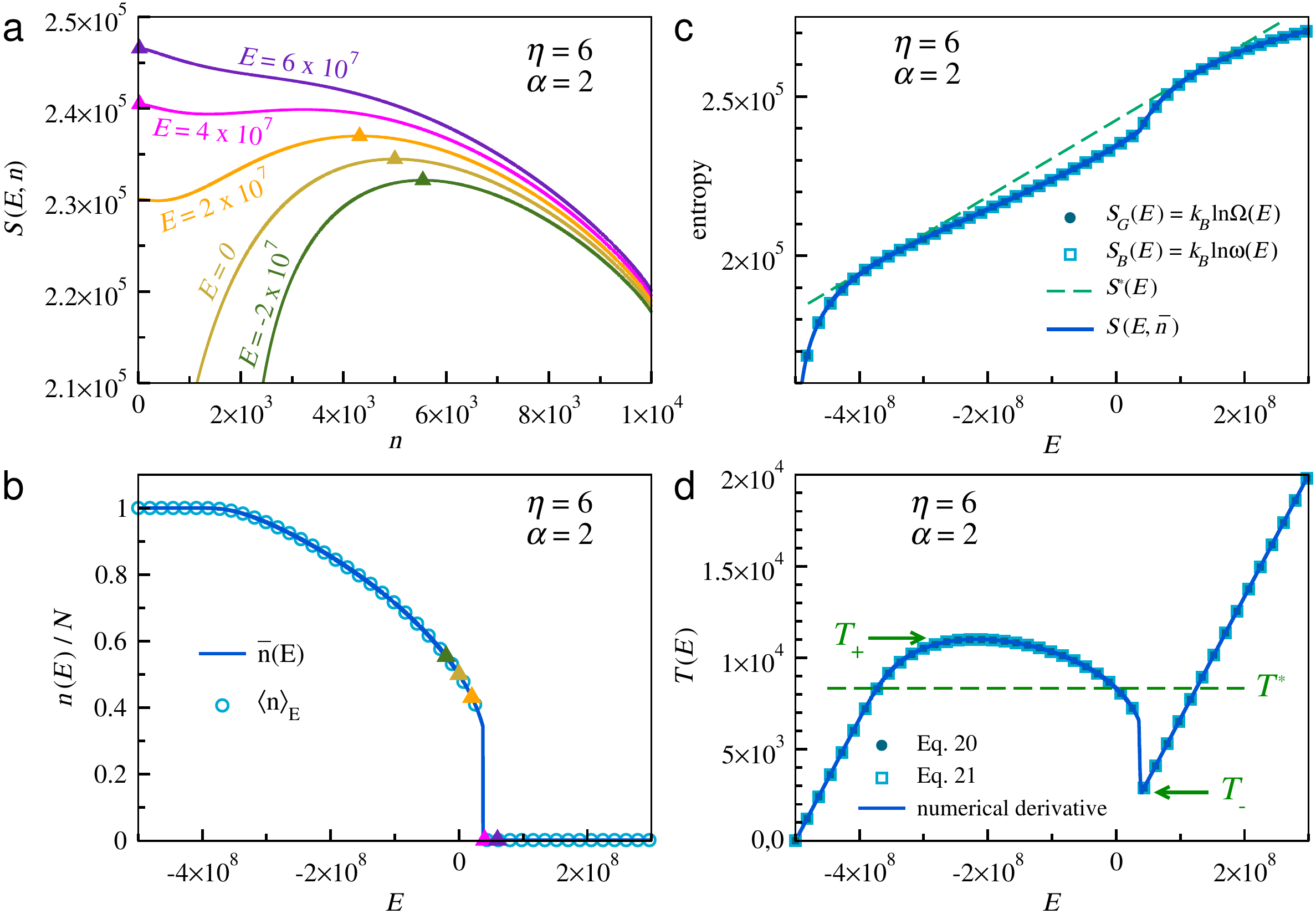}
	\caption{(a)~Function $S(E,n)$ given by Eq.~\ref{entropia_final} for different values of the total energy $E$
equally spaced between $-2\times 10^7$ and $6\times 10^7$ (in arbitrary units).
	Filled triangles denote the number of particles $\bar{n}(E)$ that maximizes $S(E,n)$.
	(b)~Ratio between the mean number of particles inside the volume $V_{0}$ and the total number of particles $N$
in the system.
	Here $\bar{n}(E)$ correspond to the values that maximizes $S(E,n)$, while
$\langle n \rangle_{E}^{\,}$ denote the microcanonical averaged values computed as in Eq.~\ref{valor_medio}.
	Filled triangles indicate the values of $\bar{n}(E)$ at the energies displayed in (a).
	(c)~Microcanonical entropies evaluated considering the Boltzmann $S_B(E)$ and Gibbs $S_G(E)$ definitions
(see Sec.~\ref{boltzmann_vs_gibbs}); $S(E)=S(E,\bar{n})$ is given by Eq.~\ref{micro_entropy_def} and corresponds 
to the maximization of Eq.~\ref{entropia_final} with respect to $n$, and $S^{*}(E)$ denote a linear 
function, Eq.~\ref{linear_entropy}, defined at the transition temperature $T^*$ (see Sec.~\ref{free_energy}).
	(d)~Microcanonical temperatures $T(E)$ obtained from a numerical derivative (continuous line) and
 using Eq.~\ref{equiparticao} (filled circles) and Eq.~\ref{equipw} (open squares). 
	The canonical transition temperature $T^* \approx 8.3\times 10^3$ is obtained through a Maxwell-like construction
 (green dashed line) and temperatures $T_- \approx 2.5 \times 10^3$ and $T_+ \approx 1.1 \times 10^4$ delimits the metastability region
where $dT(E)/dE<0$.
	Results were obtained for $\alpha=2$, $N=10^4$ particles, $\nu=5$ (in arbitrary units), 
$\eta=6$ ($\rho\approx 2.5 \times 10^{-3}$), and assuming $k_B=1$.}
	\label{fig:methods:SE}
\end{figure*}

	Figure~\ref{fig:methods:SE}(a) depicts the function $S(E,n)$ given by
Eq.~\ref{entropia_final} as a function of the number $n$ of particles inside the volume $V_0$ at different 
values of the total energy $E$.
	The filled triangles indicate the maxima of $S(E,\bar{n})$ from where one can see that, 
as the energy increases, the whole entropy curve also increases (in accordance with Fig.~\ref{fig:methods:SE}(c)). 
	In Fig.~\ref{fig:methods:SE}(b) we show the corresponding values of $\bar{n}(E)$ that maximizes
$S(E,n)$, which decrease as the total energy $E$ increases,
indicating that lower the energy, higher the number of particles in the aggregate.
	The filled triangles in Fig.~\ref{fig:methods:SE}(b) indicate
the pair $(E,\bar{n})$ in correspondence with the filled triangles that are displayed in Fig.~\ref{fig:methods:SE}(a).
	Interestingly, one can note from Fig.~\ref{fig:methods:SE}(b) that, at lower values of energy, the curve $\bar{n}(E)$ is continuous, however, at energies between the orange and magenta lines in 
Fig.~\ref{fig:methods:SE}(a), the change in the number of aggregated particles is quite abrupt, which indicate 
the presence of a first-order phase transition.
	This can be explained by the change in shape of the curves $S(E,n)$ in panel (a) as the energy increases. 
	At energies near the one represented in magenta, $S(E,n)$
has two local maxima, so that there exists an energy at which occur a change in the location 
of the global maximum, resulting in the jump in $\bar{n}(E)$ which is characteristic of
first-order phase transitions.
	It is worth noting that Fig.~\ref{fig:methods:SE}(b) also shows 
that the most probable value $\bar{n}(E)$ agrees with the microcanonical average values
$\langle n\rangle_{E}^{\,}$ evaluated as prescribed below in Sec.~\ref{micro_average}.

\subsection{Boltzmann versus Gibbs microcanonical entropies}
\label{boltzmann_vs_gibbs}

	In Sec.~\ref{entropia} we have computed the microcanonical entropy 
$S(E)$ through Eq.~\ref{micro_entropy_def} by assuming the value $\bar{n}$ which maximizes the sum in Eq.~\ref{OmegaEsum}.
	Rigorously speaking, however, it should have been evaluated by considering the whole sum.
	Because Eq.~\ref{OmegaEsum} was obtained by means of $\Omega(E)$ given by Eq.~\ref{Omega}, which accounts for the number of microscopic states with energy less than or equal to $E$, we termed the entropy computed through Eq.~\ref{OmegaEsum} as the
``Gibbs entropy'' $S_G(E)=k_B \ln \Omega(E)$.
	As illustrated in Fig.~\ref{fig:methods:SE}(c), no differences can be observed between the $S(E)=S(E,\bar{n})$ and $S_G(E)$.
	Additionally, since there are other possible definitions for the microcanonical entropy in
literature~\cite{pearson1985prA,frenkel2015amjphys,swendsen2015pre,matty2017physA}, we also assess the so-called ``Boltzmann entropy'', 
which can be defined as $S_B(E)=k_B \ln \omega(E)$, where $\omega(E)$ is the number of microscopic states in the vicinity of $E$, {\it i.e.}, between $E$ and $E+\varepsilon$, with $\varepsilon$ being a small energy interval (see, {\it e.g.}, Ref.~\cite{kubobook}).
	In the case of classical systems, $\omega(E)$ is obtained by integrating the phase space only in the vicinity of $E$, which is done by replacing the step function in Eq.~\ref{Omega} by a delta 
$\delta(E-H(\textbf{q},\textbf{p}))$.
	Since the potential energy does not depend on the momentum coordinates, 
we get the analogue of Eq.~\ref{Omega_q}
\begin{equation}
\omega(E)=\left(\frac{2\pi m}{h^2}\right)^{3N/2} \frac{\varepsilon}{N! \, \Gamma(3N/2)}
\int\text{d}\textbf{q}\,(E-E_p(\textbf{q}))^{3N/2-1} \, \Theta(E-E_p(\textbf{q})).
\label{omega_qA}
\end{equation}
	Thus, for the generalized aggregation model considered here, in particular,
one finds an expression which is similar to Eq.~\ref{Omega_E}, that is,
\begin{equation}
\omega(E)\propto\sum_{n=n_{min}}^{N} \Omega_p(n)(E-E_p(n))^{3N/2-1},
\label{omega_EA}
\end{equation}
with $\Omega_p(n)$ given by Eq.~\ref{densidade_eta}.
	Figure~\ref{fig:methods:SE}(c) indicates that, for the model considered here with $N=10^4$ particles, both microcanonical entropies $S_G(E)$ and $S_B(E)$ defined, respectively, from Eqs.~\ref{Omega_E} and~\ref{omega_EA}, yield exactly the same behaviour.
	Importantly, the results presented in Fig.~\ref{fig:methods:SE}(c) also indicate that $S_B(E)$, just as $S_G(E)$, agrees with the microcanonical entropy obtained by maximization of Eq.~\ref{entropia_final} with respect to $n$, {\it i.e.}, $S(E)=S(E,\bar{n})$.

\subsection{Microcanonical average values}
\label{micro_average}

	In general, by considering the microcanonical ensemble of a classical system defined with fixed energy $E$, 
one can evaluate the mean value of a generic quantity $A(\textbf{q},\textbf p))$ that is a function of the positions
and momentum of all particles $(\textbf{q},\textbf p)$ through the sum, in phase space, of all possible and equally 
probables values of that quantity at the vicinity of $E$, divided by the total number microscopic states around 
that energy, $\omega(E)$, that is~\cite{pearson1985prA}
\begin{equation}
\langle A\rangle_E=\frac{\int\text{d}\textbf{q} \, \text{d}\textbf{p}  \,  A(\textbf{q},\textbf{p})  \,  \delta(E-H(\textbf{q},\textbf{p}))} {\int\text{d}\textbf{q}  \,  \text{d}\textbf{p}  \, \delta(E-H(\textbf{q},\textbf{p}))},
\label{mean_value}
\end{equation}
where the constant $\varepsilon/N!  \, h^{3N}$ cancels out. 
	Again, by assuming that the potential energy is independent of the momentum coordinates and the quantity $A$ is a function only of the position coordinates, the momentum part can be integrated in both the numerator and denominator of Eq.~\ref{mean_value}. 
	In the case of our generalized model, the variable that carries the information about the position of particles is the number $n$ of particles in the aggregate, so that one can get an expression for the mean values of a quantity $A(n)$ as
\begin{equation}
\langle A\rangle_{E}^{\,}=\frac{\sum_{n=n_{\min}}^{N}\Omega_p(n)\, A(n) \,(E-E_p(n))^{3N/2-1}} {\sum_{n=n_{\min}}^{N}
\Omega_p(n) \, (E-E_p(n))^{3N/2-1}}.
\label{valor_medio}
\end{equation} 
	By setting $A(n)=n$ in the above equation one gets the expression for $\langle n\rangle_{E}^{\,}$ 
that is shown, for instance, in Fig.~\ref{fig:methods:SE}(b).
	It is worth mentioning that there is a very good agreement between $\langle n\rangle_{E}^{\,}$ and 
the number of particles in the aggregate, $\bar{n}(E)$, which is obtained from the maximization of the function $S(E,n)$ defined by Eq.~\ref{entropia_final}.

\subsection{Microcanonical temperatures}

	Usually, one can consider the definition based on thermodynamics to compute 
the microcanonical temperature $T(E)$ as~\cite{kubobook}
\begin{equation}
T(E)=\left(\frac{\partial S}{\partial E}\right)_{N,V}^{-1}.
\label{temperature}
\end{equation}
	Thus, $T(E)$ can be readily estimated from microcanonical entropy by means of a numerical derivative.
	Although such straightforward method seems to be very simple and practical, it might lead to numerical inaccuracies, mainly due to the lack of a systematic way of choosing the number of points considered in the numerical derivative. 
	Alternatively, it is possible to derive an expression based on the averaging procedure discussed in Sec.~\ref{micro_average}.
	The important step of the derivation is to write the density of states $\Omega(E)$ in terms of the mean value of kinetic energy $\langle E_k\rangle_{E}^{\,}$ computed as in Eq.~\ref{mean_value}.
	Hence, by noting that kinetic energy can be expressed as a function of the positions coordinates, $E_k(\textbf{q})=E-E_p(\textbf{q})$, because total energy $E$ is fixed, and carrying out the derivatives in Eq.~\ref{temperature}, one gets~\cite{pearson1985prA,schierz2015jcp}
\begin{equation}
T(E)=\frac{2k_B^{-1}}{3N}\langle E_k\rangle_{E}^{\,},
\label{equiparticao}
\end{equation} 
which is equivalent to the equipartition theorem. 
	It is worth noting that Eq.~\ref{equiparticao} was obtained using the Gibbs definition of entropy $S_G(E)=k_B\ln 
\Omega(E)$.
	Alternatively, by choosing the Boltzmann definition $S_B(E)=k_B\ln\omega(E)$, one finds
a slightly different expression~\cite{pearson1985prA,schierz2015jcp},
\begin{equation}
T(E)=\frac{2k_B^{-1}}{3N-2}\frac{1}{\langle E_k^{-1}\rangle_{E}^{\,}}.
\label{equipw}
\end{equation}
	In Fig.~\ref{fig:methods:SE}(d) we illustrate the microcanonical temperatures 
evaluated via the three methods in a region where $T(E)$ display a S-shape behaviour which is typical of first-order phase transitions~\cite{schnabel2011pre}.
	As it can be seen in Fig.~\ref{fig:methods:SE}(d), the agreement between Eqs.~\ref{equiparticao} and \ref{equipw} is already obtained in the case of $N=10^4$ particles.
	Besides, as shown in that figure, the numerical derivative of the microcanonical entropy $S(E)$ given by Eq.~\ref{micro_entropy_def} also yields a very accurate microcanonical temperature.
	The equivalence between the entropy definitions, {\it i.e.}, $S(E)$, $S_G(E)$, and $S_B(E)$, as well as the agreement between the most probable value $\bar{n}(E)$ and mean value $\langle n\rangle_{E}^{\,}$ are complementary, and occurs because, although finite, the value of $N$ is already large enough.

\subsection{Free-energy profiles and free-energy barriers}
\label{free_energy}

Until here our analysis was entirely performed in the microcanonical ensemble, but, from $S(E)$, it is straightforward to obtain the canonical probability distributions $p(E;\beta)$ given at a fixed temperature $T=1/k_B\beta$, {\it i.e.},
\begin{equation}
p(E;\beta)\propto 
\omega(E)e^{-\beta E} \, \approx \, e^{-\beta(E-TS(E))}\equiv 
e^{-\beta F(E;\beta)},
\end{equation} 
with $F(E;\beta)$ being an energy dependent Helmholtz-like free energy defined as $F(E;T)\equiv E-TS(E)$ (see, {\it e.g.}, Ref.~\cite{matty2017physA}). 
	If the system presents a first-order phase transition and it is in contact with a thermal reservoir at a temperature that is close to the transition temperature, the canonical distribution $p(E;\beta)$ should display two maxima at energies $E_-$ and $E_+$ separated by a minimum at energy $E^*$, and those correspond to the two minima and one maximum of the free energy $F(E;T)$, respectively.
	The two minima in the free energy $F(E;T)$ indicate the presence of two 
stable states~\cite{kosterlitz1990prl}.
	At the transition temperature, $T^*=1/k_B\beta^*$, the corresponding canonical distribution 
$p^*(E)\equiv p(E;\beta^*)\propto e^{-\beta^*\Delta F(E)}$ has its two maxima with equal heights, and one can define a
free-energy profile as
\begin{equation}
\beta^*\Delta F(E)\equiv\beta^*[F(E;T^*)-F(E_-;T^*)],
\label{dF_e}
\end{equation}
so that the two minima of the free-energy profile lie in the energy-axis at energies $E_-$ and $E_+$ (see Fig.~\ref{fig:conformational}(a) for an example of a free-energy profile).

	As discussed in Refs.~\cite{janke1998nuclphys,janke2017natcommun,rizzi2020jstat}, the criteria of equal height for the maxima of the canonical distribution is equivalent to the Maxwell-like construction indicated in Fig.~\ref{fig:methods:SE}(d), where the horizontal dashed (green) line represent the temperature $T^*$ that delimits equal areas, above and below, of the microcanonical temperature curve $T(E)$.
	The Maxwell-like construction also determines the values for the energies $E_-$
and $E_+$ which are used to evaluate the linear function 
\begin{equation}
S^{*}(E)= k_B \beta^{*}(E-E_-) + S(E_-),
\label{linear_entropy}
\end{equation}
that is displayed in Fig.~\ref{fig:methods:SE}(c).
	This linear function is defined in way that $S^{*}(E_-)=S(E_-)$
and $S^{*}(E_+)=S(E_+)$, so that, since $\beta^{*}=1/k_BT^{*}$, it allows one to 
determined the inverse of the transition temperature  consistently as
\begin{equation}
\frac{1}{T^{*}} = \frac{S(E_+) - S(E_-)}{\Delta E^{\dagger}},
\end{equation}
with $\Delta E^{\dagger}=E_+ - E_-$ being an estimate for the the latent heat.

	Additionally, from the free-energy profile $\beta^*\Delta F(E)$,
it is straightforward to compute the free-energy barrier $\beta^*\Delta F^\dagger\equiv\beta^*\Delta F(E^*)$, which is the barrier that has to be surpassed by the system in order to assemble or disassemble an aggregate at the transition temperature $T^{*}$.
	As illustrated in Fig.~\ref{fig:conformational}(a), the energy $E^*$, as well as the barrier height $\beta^*\Delta F^\dagger$, can be directly identified from the maximum of the free-energy profile.

	Finally, it is worth noting that, if the temperature $T$
of the thermal reservoir is far from $T^*$, the canonical distribution $p(E;\beta)$ will present a single maximum.
	As indicated in Fig.~\ref{fig:methods:SE}(d), the microcanonical temperature curve $T(E)$ allows one to identify 
the limiting temperatures $T_-$ and $T_+$, which delimits the region of thermodynamic instability/metastability where $dT/dE<0$, so that, only if the temperature $T$ is within this range, the system will
have non-zero probabilities to be found in either one of the two phases (see, {\it e.g.}, Ref.~\cite{frigori2010jphysconfser}).

\subsection{Conformational microcanonical ensemble}
\label{conformational}

	Even though it is possible to integrate 
the momentum coordinates
in many physical systems~\cite{janke2017natcommun,janke2017jphysconfser,schierz2016pre},
	it has been a common practice in computational studies (see, {\it e.g.}, Refs.~\cite{pleimling2001jstatphys,pleimling2005pt,beath2006prb,behringer2006pre,martinmayor2007prl,nogawa2011pre,rizzi2016prl,rizzi2011jcis,junghans2006prl,chen2008pre,junghans2008jcp,junghans2009epl,trugilho2020jphysconfser,church2012jcp,taylor2009pre,taylor2009jcp,taylor2010physproc,scheraga1994jphyschem,chen2007pre,rojas2008prl,bereau2010jacs,liu2012jcp,frigori2013jcp,frigori2014pre,alves2015cpc,frigori2017pccp,frigori2021jmolmod,chen2009jcp,wangliang2009jcp,moddel2010pccp}) 
to perform the microcanonical analysis directly from the conformational microcanonical ensemble, {\it i.e.}, neglecting the kinetic energy contribution to $E$, thus
performing the analysis based only on the conformational density of states $\Omega_p(E_p)$.
	The conformational entropy of a system with fixed potential energy $E_p$ is then defined as~\cite{schierz2016pre}
 $S_p(E_p)=k_B\ln \Omega_p(E_p)$.
	This kind of approach looks convenient mainly when working with Monte Carlo simulations due to the existence of
 algorithms and techniques that provide direct access to $\Omega_p(E_p)$, {\it e.g.},
multicanonical~\cite{berg1992prl,berg2003cpc}, 
entropic sampling~\cite{lee1993prl},
broad histogram~\cite{broadhist1996brazjphys},
Wang-Landau~\cite{wanglandau2001prl},
and 
statistical temperature~\cite{straub2011jcp,rizzi2011jcp}.
	Even so, it is not obvious that the kinetic contribution can be neglected in the case of finite-sized systems with aggregating particles in contact with a thermal reservoir.

\begin{figure*}[!t]
\centering
\includegraphics[width=0.95\textwidth]{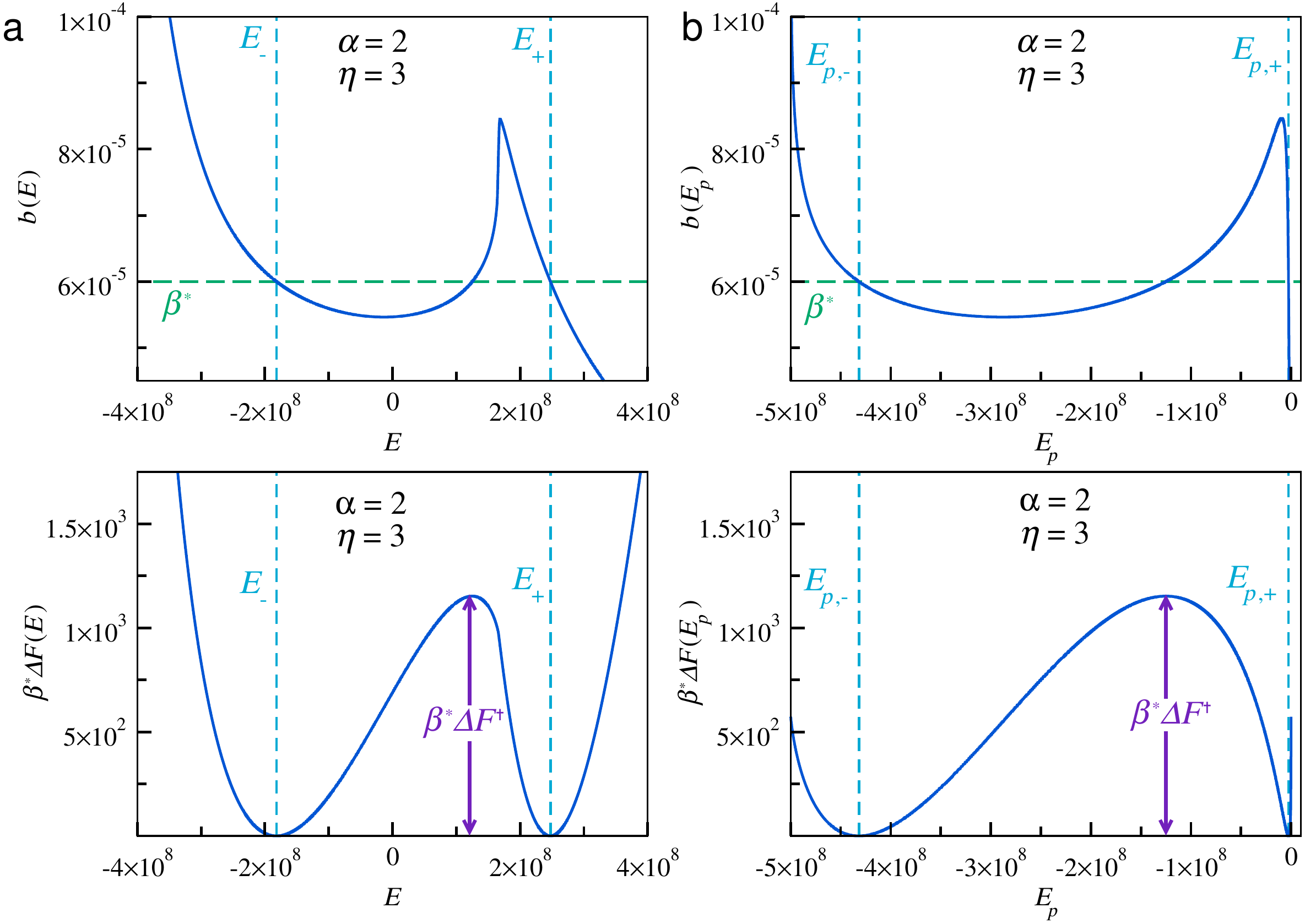}
\caption{(a)~Inverse of the microcanonical temperature $b(E)=1/T(E)$ and free-energy profile $\beta^*\Delta F(E)$ as functions of the total energy $E$. 
	(b)~Inverse of the microcanonical temperature $b(E_p)=1/T(E_p)$ and free-energy profile 
$\beta^*\Delta F(E_p)$ as functions of the potential energy $E_p$. 
	Horizontal dashed (green) lines denote the inverse of transitions temperatures $\beta^{*}=1/T^{*} \approx 6 \times 10^{-5}$ that were obtained via Maxwell-like constructions. 
	Vertical dashed (blue) lines indicate the minima of the free-energy profiles, from where one can extract the latent heats $\Delta E^\dagger=E_+-E_-$ and $\Delta E_p^\dagger=E_{p,+}-E_{p,-}$.
	As indicated by the arrows, free-energy barriers, $\beta^*\Delta F^\dagger$, are evaluated from the difference between the maximum and the minima of $\beta^*\Delta F(E)$.
	Results were obtained for $\alpha=2$, $\eta=3$, $N=10^4$, $\nu=5$ (in arbitrary units), and assuming $k_B=1$.
}
\label{fig:conformational}
\end{figure*}

	Since we have already obtained $\Omega_p(n)$ for our aggregation model, {\it i.e.}, Eq.~\ref{densidade_eta}, 
the conformational entropy $S_p(n)=k_B\ln \Omega_p(n)$ can be evaluated through the already used Stirling's approximation, that is,
\begin{equation}
S_p(n)\approx\eta(N-n)-\frac{1}{2}\ln(1+2\pi n)-n\ln n \\
-\frac{1}{2}\ln(1+2\pi(N-n)) -(N-n)\ln(N-n),
\label{S_Ep}
\end{equation}
where we have omitted the additive constants which depend on $N$ and $\eta$, but not those that depend on $n$, and set $k_B=1$ to abbreviate the notation.
	Besides, by recalling that $E_p(n)=-\nu g(n)$ with $g(n)$ given by Eq.~\ref{gn_alpha},
one can invert that expression in order to obtain the number of particles $n$ inside the volume $V_0$ in terms of the potential energy, that is,
\begin{equation}
n(E_p)=\left(1-\frac{E_p}{\nu}\right)^{1/\alpha},
\label{n_Ep}
\end{equation}
so that the conformational entropy given by Eq.~\ref{S_Ep} can be rewritten in terms of $E_p$,
{\it i.e.}, $S_p(E_p)$.
	Hence, in analogy to Eq.~\ref{temperature}, the microcanonical temperature in the conformational ensemble
can be computed as
\begin{equation}
T(E_p)=\left(\frac{\partial S_p}{\partial E_p}\right)_{N,V}^{-1}.
\end{equation}

	Here, just as in the case of the usual microcanonical ensemble where one has that
$b(E)=1/T(E)$, one can use Eqs.~\ref{S_Ep} and~\ref{n_Ep} to obtain an expression for the inverse 
of temperature $b(E_p)=1/T(E_p)$, that is,
\begin{eqnarray}
b(E_p)= \left[\eta +\frac{\pi}{1+2\pi n(E_p)} -\frac{\pi}{1+2\pi(N-n(E_p))} -\ln\left(\frac{N-n(E_p)}{n(E_p)}\right)\right] \nonumber \\ 
~~ ~~ ~~ ~~ ~~ ~~ ~~ ~~ ~~ ~~ ~~ ~~ 
\times \left(\frac{1}{\alpha\nu}\right) \left(1-\frac{E_p}{\nu}\right)^{(1-\alpha)/\alpha},
\label{b_Ep}
\end{eqnarray}
with $n(E_p)$ given by Eq.~\ref{n_Ep}.

	In Fig.~\ref{fig:conformational} we show curves for the inverse of microcanonical temperature and free-energy profiles obtained from both the usual microcanonical and the conformational ensemble approaches.
	While $b(E)$ in Fig.~\ref{fig:conformational}(a) is obtained through the numerical derivative of $S(E)$ with respect to the total energy (see Eq.~\ref{temperature} in Sec.~\ref{entropia}), $b(E_p)$ in Fig.~\ref{fig:conformational}(b) is given by Eq.~\ref{b_Ep}.
	In both cases, the transition temperatures $T^*=1/\beta^*$ and free energy-profiles $\beta^{*}\Delta F(E)$ are 
obtained via the Maxwell-like construction discussed in Sec.~\ref{free_energy}
	(noting that similar analyses and definitions can be applied to the case of the conformational ensemble).
	Thus, it is straightforward to compute energies $E_{p,-}$, $E_{p,+}$, and $E_{p}^{*}$, from where
one can determine latent heats and free-energy barriers as defined in Sec. \ref{free_energy}.
	Figure~\ref{fig:conformational} indicates that, although there are
clear differences between the curves $b(E)$ and $b(E_p)$ as well as between 
the free-energy profiles $\beta^{*}\Delta F(E)$, both approaches yield the same transition temperatures, $T^{*}=1/\beta^{*} \approx  1.67 \times 10^4$,
latent heats, 
$\Delta E^{\dagger}=\Delta E_p^{\dagger} \approx 4.3 \times 10^8$,
and free-energy barriers, $\beta^{*}\Delta F^{\dagger} \approx 1.15 \times 10^3$.
	As we will show in Secs.~\ref{phase_diagram} and~\ref{barrier}, such agreement is also observed for other values
 of the parameters $\alpha$ and $\eta$.

\section{Results}

\subsection{Case $\alpha=2$}
\label{alpha2}

\begin{figure*}[!t]
\centering
\includegraphics[width=0.95\textwidth]{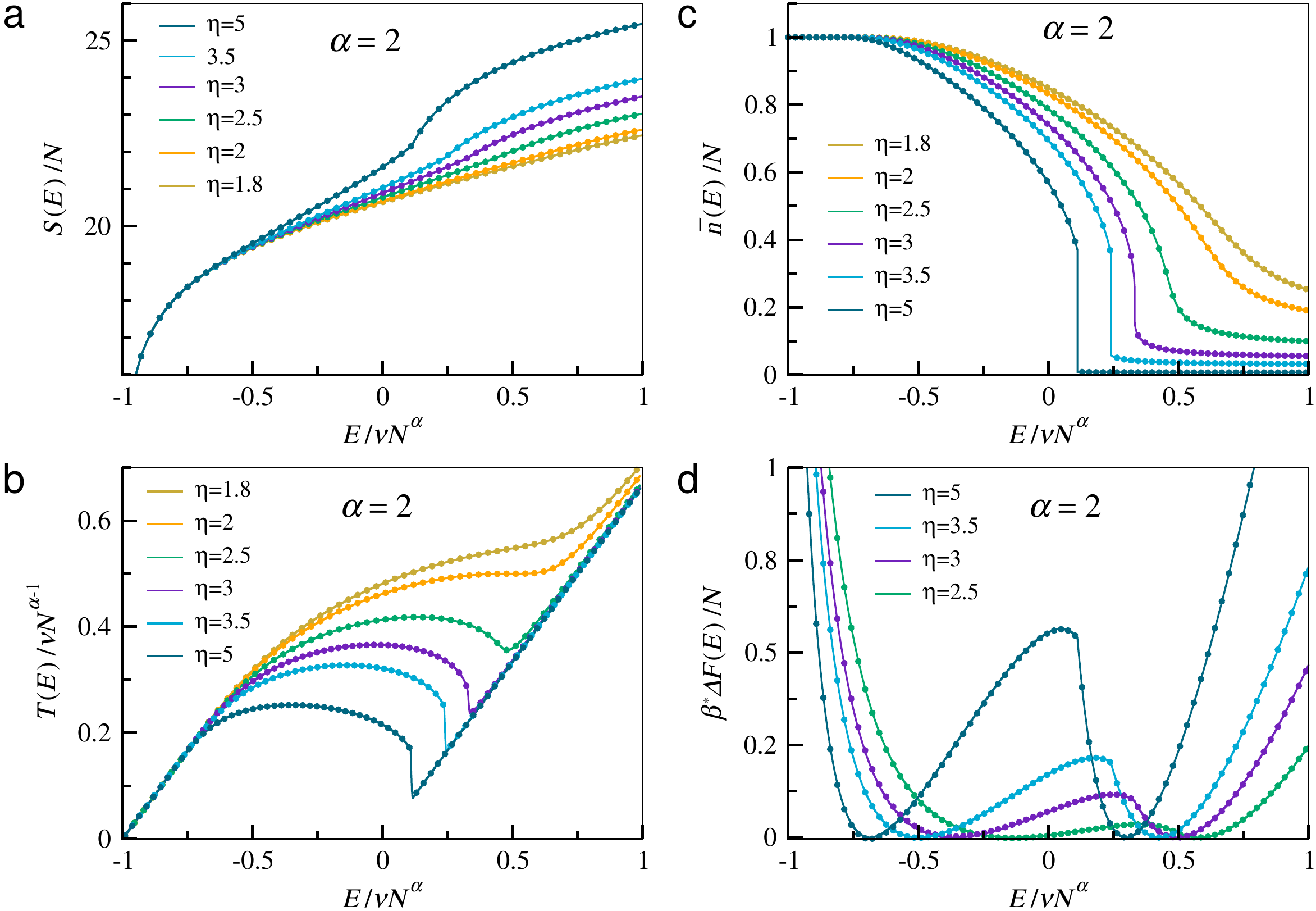}
\caption{
(a)~Microcanonical entropy per particle, $S(E)/N$, as a function of the rescaled energy $E/\nu N^{\alpha}$ obtained for $\alpha=2$ and different $\eta$. 
(b)~Rescaled microcanonical temperature, $T(E)/\nu N^{\alpha-1}$.
(c)~Average fraction of the number of particles  in the aggregate, $\bar{n}(E)/N$. 
(d)~Free-energy profile at the transition temperature, $\beta^{*}\Delta F(E)$. 
Continuous lines corresponds to data obtained for $N=5 \times 10^3$ particles and $\nu=5$, 
while filled circles indicate the results obtained for $N=10^4$ particles and $\nu=9$. 
All results are displayed assuming $k_B=1$.
}
\label{fig:alpha2}
\end{figure*}

	In Fig.~\ref{fig:alpha2} we present the results obtained for our generalized model
with $\alpha=2$ for different values of the volume parameter $\eta$.
	As mentioned earlier, this value of $\alpha$ corresponds to the original Thirring's 
model where it is assumed that all the $n$ particles inside the volume $V_0$ 
interact with each other through a mean-field-like long range interaction.

	As previously indicated in Ref.~\cite{campa2016jstat} for $\alpha=2$, the total energy and the average number of particles inside $V_0$ scale to the number of particles $N$ and the magnitude of interaction $\nu$ as $E \propto \nu N^{2}$ and $\bar{n}(E) \propto N$.
	In order to verify that and other scaling relations, we include in Fig.~\ref{fig:alpha2} the rescaled data for $N=5\times10^3$ and $\nu=5$ (continuous lines), and $N=10^4$ and $\nu=9$ (filled circles).
	As it can be seen from the data collapse in Fig.~\ref{fig:alpha2}(a), although the system
is indeed non-additive~\cite{campa2016jstat}, and the energy scales as $E\propto\nu N^\alpha$ 
with $\alpha=2$, the microcanonical entropy scales as $S(E)\propto N$.
	As a consequence, the microcanonical temperature $T(E)$ evaluated from $S(E)$ through Eq.~\ref{temperature} will not correspond to an intensive quantity since it will behave like $T(E)\propto\nu N^{\alpha-1}$.
	Figure~\ref{fig:alpha2}(b) indicate that this is indeed valid, at least for $\alpha=2$, so that $T(E)\propto\nu N$.
	In addition to the expected linear behaviour for the averaged number of 
particles in the aggregate, {\it i.e.}, $\bar{n}(E)\propto N$,  displayed in Fig.~\ref{fig:alpha2}(c), Fig.~\ref{fig:alpha2}(d) indicate that the free-energy profile, which is defined in terms of energy per temperature, also scales linearly with the total number of particles in the system, {\it i.e.}, $\beta^*\Delta F(E)\propto N$.

	In addition to the aforementioned scaling behaviours, it is also important to discuss the thermostatistics of the model for the different values of the ``reduced'' volume $\eta$.
	By recalling that $\eta$ is related to the volumes $V$ and $V_0$ through Eq.~\ref{eta}, 
one might interpret higher values of $\eta$ as higher ratios between the 
total volume $V$ and the volume $V_0$.
	In Fig.~\ref{fig:alpha2}(a), in particular, where we show the microcanonical entropy per particle $S(E)/N$ as a function of the rescaled energy $E/\nu N^\alpha$, one can see that for higher values of $\eta$, there are regions where 
$S(E)$ is a convex function of energy, so that $\left( \partial^2 S/\partial E^2 \right)_{N,V}>0$.
	This behaviour can be better identified from Fig.~\ref{fig:alpha2}(b), where it is shown the rescaled microcanonical temperature $T(E)/\nu N^{\alpha-1}$ as a function of rescaled energy.
	It is clear from the S-shaped curves observed for $\eta\geq 2.5$ that there are regions where the temperature  $T(E)$ decreases with energy, indicating the presence of a ``convex intruder'' in $S(E)$~\cite{grossbook,schnabel2011pre}.
	The data in Fig.~\ref{fig:alpha2}(c) for the average
fraction of particles in the aggregate, $\bar{n}(E)/N$, clear indicate 
that there is an aggregation transition from a highly energetic diluted phase 
to a phase where all the particles should be inside $V_0$ at low energies.
	Here it is worth noting that a transition temperature $T^{*}$ will only be defined through a Maxwell-like construction if the microcanonical temperature $T(E)$ presents a S-shaped curve with a region where it decreases with energy (see Sec.~\ref{free_energy} for details).
	As a consequence, the free-energy profiles $\beta^{*}\Delta F(E)$ will only present two different minima
at this condition, {\it i.e.}, for $\eta\geq 2.5$, just as illustrated in Fig.~\ref{fig:alpha2}(d).
	In general, the presence of a free-energy barrier as well as of a latent heat can be used to indicate the presence of first-order phase transitions
in the canonical ensemble~\cite{schnabel2011pre}.

	Here it is worth mentioning that, rigorously, because the system is intrinsically finite and non-additive ({\it i.e.}, display long range interactions for $\alpha=2$), the canonical critical point at $\eta_{\text{cp}}^{\,}=2$ might be different from its microcanonical counterpart (see Refs.~\cite{campa2009physrep,campa2016jstat}).
	This means that, although the system present a first-order phase transition in the canonical ensemble 
for $\eta = 2.5$ (which is confirmed by the both, {\it e.g.}, the S-shaped temperatures $T(E)$ and the presence 
of two minima in $\beta^{*} \Delta F(E)$), the microcanonical curve $\bar{n}(E)$ displayed in Fig.~\ref{fig:alpha2}(c) 
seems to be continuous (see Ref.~\cite{campa2016jstat} for further details).
	Even so, for $\eta \geq 3$ both canonical and microcanonical ensembles are fully equivalent with respect to 
the classification of the nature of the aggregation transition.

\begin{figure*}[!b]
\centering
\includegraphics[width=0.95\textwidth]{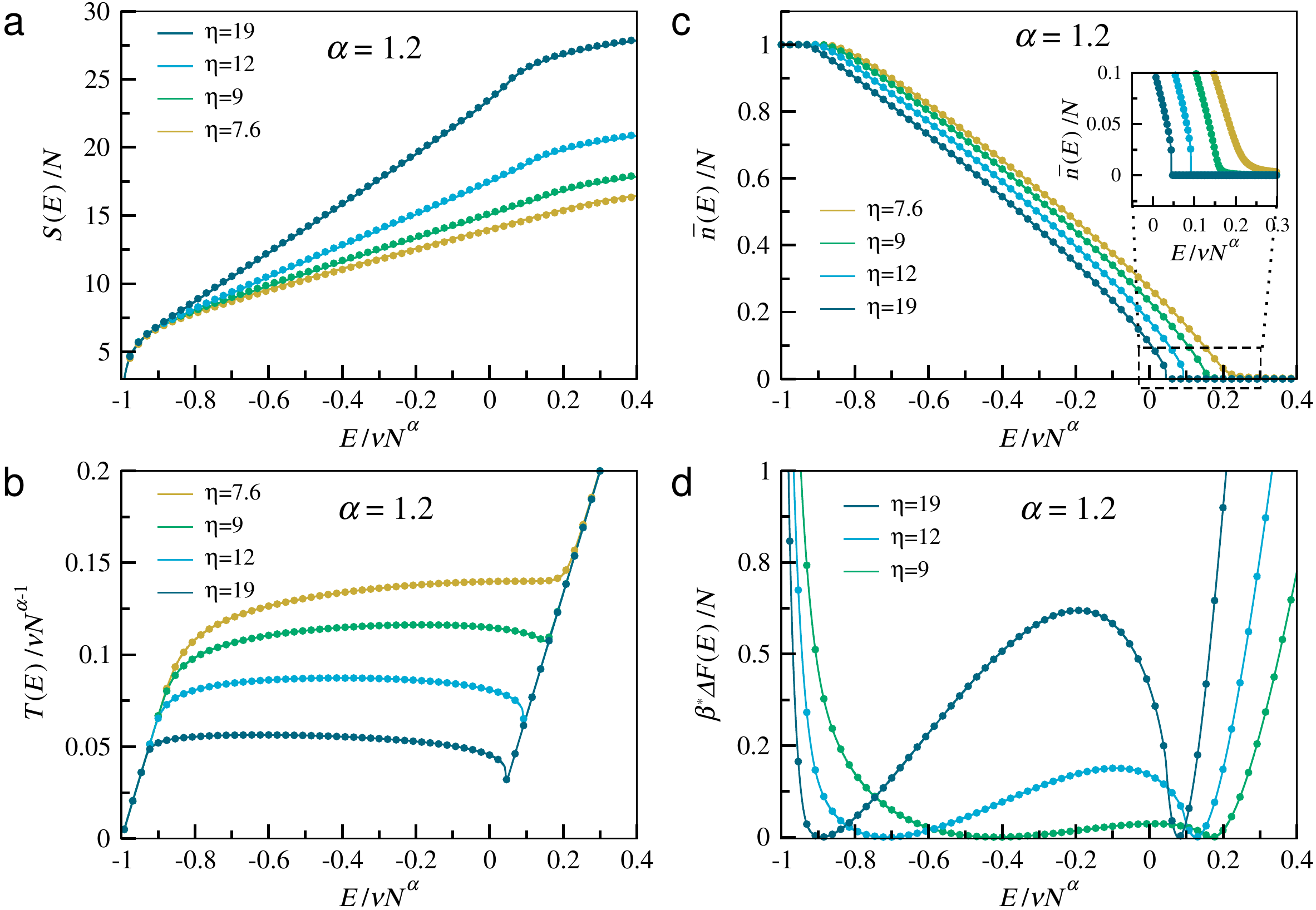}
\caption{
(a)~Microcanonical entropy per particle, $S(E)/N$, as a function of the rescaled energy, $E/\nu N^{\alpha}$ obtained for $\alpha=1.2$ and different $\eta$. 
(b)~Rescaled microcanonical temperature $T(E)/\nu N^{\alpha-1}$.
(c)~Average fraction of the number of particles in the aggregate, $\bar{n}(E)/N$.
(d)~Free-energy profile at the transition temperature, $\beta^{*}\Delta F(E)$.
Continuous lines corresponds to data obtained for $N=5 \times 10^3$ particles and $\nu=5$,
while filled circles indicate the results obtained for $N=10^4$ particles and $\nu=9$.
All results are displayed assuming $k_B=1$.}
\label{fig:alpha12}
\end{figure*}

\subsection{Case $\alpha=1.2$}

	Next, in Fig.~\ref{fig:alpha12}, we present a similar thermostatistics analysis but now for a potential energy $E_p(n)$ defined by an effective number of bonds (Eq.~\ref{gn_alpha}) computed with $\alpha=1.2$.
	As mentioned earlier, that value of $\alpha$ is used in order to describe finite-sized disordered aggregates where particles at its surface contribute differently from those 
which are in its centre. 
	Just as in the previous section, the rescaled data for $N=5\times10^3$ and $\nu=5$ are represented by the continuous lines, while the rescaled data for $N=10^4$ and $\nu=9$ correspond to the filled circles.
	Importantly, Fig.~\ref{fig:alpha12} indicates that all quantities scale in the same way as discussed in Sec.~\ref{alpha2}, {\it i.e.}, $E \propto \nu N^{\alpha}$, $S(E) \propto N$, $T(E) \propto \nu N^{\alpha-1}$, $\bar{n}(E) \propto N$, 
and $\beta^{*}\Delta F(E) \propto N$.
	Also, the general qualitative behaviour is not significantly altered, that is, for higher values of the ``reduced'' volume ({\it i.e.}, for $\eta \geq 9$), one can observe regions where the microcanonical entropy $S(E)$ is a convex function of energy (see Fig.~\ref{fig:alpha12}(a)).
	Accordingly, this behaviour, which is consistent with first-order phase transitions in the canonical ensemble, is also corroborated by the fact that there are regions where the microcanonical temperatures $T(E)$ decrease with energy, Fig.~\ref{fig:alpha12}(b), and the free-energy profiles $\beta^{*}\Delta F(E)$ present two minima, as in Fig.~\ref{fig:alpha12}(d).
	From the inset of Fig.~\ref{fig:alpha12}(c) one can see that (at least for the values considered here) the change in 
 $\bar{n}(E)/N$ seems to be discontinuous only for $\eta\geq 12$.
	It is worth noting that the results displayed in Fig.~\ref{fig:alpha12} for $\alpha=1.2$ present a behaviour that is qualitatively similar to what is observed for other aggregation models~\cite{mueller2015physproc,zierenberg2014jcp,schierz2016pre,janke2017natcommun,janke2017jphysconfser,trugilho2020jphysconfser}.

\subsection{Phase diagrams}
\label{phase_diagram}

	Although useful, the reduced ``volume'' parameter $\eta$ was introduced in Eq.~\ref{eta} only to simplify the expression for the density of states. 
	Experimentally, however, it could be more convenient to interpret the results in terms of other 
related quantities, like the total concentration of particles $C=N/V$.
	Even though there are no excluded volume interactions between point-like particles, 
one can obtain a simple relation between $C$ and $\eta$ by assuming an approximation that is
 commonly used in the context of molecular systems, which is that a particle in the aggregate
phase occupies an arbitrarily chosen volume given by $v_m$.
	Although the specific value of $v_m$ is not essential here, it may be associated to a 
molecular number density ($\sim 1/v_m$) in order to set the experimentally relevant length scales
and concentrations (see, {\it e.g.}, Ref.~\cite{trugilho2021arxiv}).
	Thus, we assume that $V_0=Nv_m$ and, in order to obtain the relation between $C$ and $\eta$, 
we use Eq.~\ref{eta} to define a third quantity $\rho$ as
\begin{equation}
\rho=\frac{V_0}{V}=\frac{1}{1+e^\eta},
\end{equation}
so that
\begin{equation}
C=\frac{N}{V}
\times\frac{V_0}{V_0}=\frac{N}{V_0}\times\frac{V_0}{V}
=\frac{\rho}{v_m},
\label{concentracao}
\end{equation}
{\it i.e.}, the concentration $C$ is directly proportional to the parameter $\rho$,
which is then considered as a dimensionless concentration.

\begin{figure*}[!t]
\centering
\includegraphics[width=0.54\textwidth]{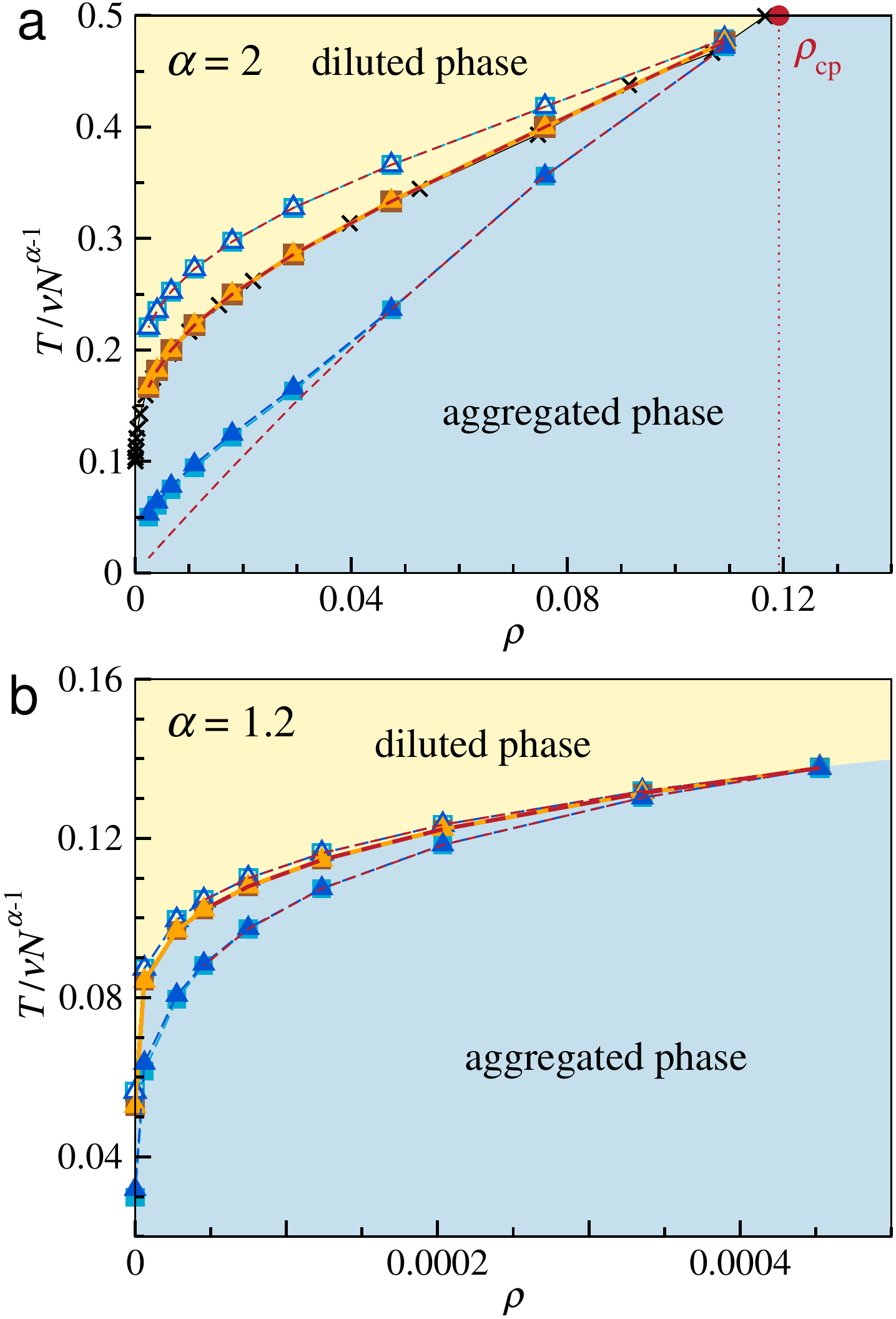}
\caption{(a) and (b) show the (rescaled) temperature 
versus concentration 
phase diagrams for  $\alpha=2$ and $\alpha=1.2$, respectively.
The continuous (orange and brown) lines represent the rescaled transition 
temperatures $T^*$, and the dashed lines indicate the rescaled temperatures $T_+$ (open symbols) and $T_-$ (filled symbols).
The open and filled squares correspond to results obtained for $\nu=5$ and $N=5\times 10^3$, 
while the open and filled triangles correspond to $\nu=9$ and $N=10^4$.
Short-dashed (red) lines denote values estimated from the approach based on the conformational microcanonical entropy.
Black crosses and the filled (red) circle/vertical dotted line in (a) are the values determined from the data presented in Ref.~\cite{campa2016jstat}.
All results are displayed assuming $k_B=1$.
}
\label{fig:phasediagram}
\end{figure*}

	In Fig.~\ref{fig:phasediagram} we include the phase diagrams for both $\alpha=2$ and $\alpha=1.2$, showing the temperatures $T^{*}$, $T_{-}$, and $T_{+}$ (see Sec.~\ref{free_energy}), all as functions of $\rho$ instead of $\eta$.
	The continuous (orange) lines represent the canonical transition temperature $T^*$, {\it i.e.}, the 
line where the system has non-zero probabilities to be found either in the diluted or in the aggregated phase.
	The dashed lines above and below $T^*$ correspond to the temperatures $T_+$ and $T_-$, respectively, which delimit the metastability region, where the canonical distribution $p(E;\beta)$ still presents two maxima and rare events such as nucleation can occur.
	Importantly, the temperature $T_-$ can be interpreted as a spinodal temperature, while
$T_+$ can be considered a solubility line, since no aggregates will be present in the system
for temperatures $T>T_{+}$.

	In the two panels of Fig.~\ref{fig:phasediagram} we include the rescaled data for different values of $N$ and $\nu$.
	The data collapse for these sets of parameters confirms that the temperatures $T^{*}$, $T_{+}$, and $T_{-}$, scale like $T\propto\nu N^{\alpha-1}$, just like it the microcanonical temperatures $T(E)$ displayed in Figs.~\ref{fig:alpha2}(b) and~\ref{fig:alpha12}(b). 
	Besides, the results for $\alpha=2$ displayed in Fig.~\ref{fig:phasediagram}(a) indicate that, except from a small deviation in $T_-$ for the lower values of $\rho$, the phase diagram obtained via the conformational microcanonical ensemble, {\it i.e.}, ignoring the kinetic contribution (see Sec.~\ref{conformational}), is in good agreement with the phase diagram obtained from the usual microcanonical ensemble.
	Even so, we note that, for $\alpha=1.2$ and lower values of $\rho$, it was not possible to perform reliable Maxwell-like constructions since the S-shape behaviour, and thus $T_{-}$, in the resulting microcanonical temperatures $T(E_p)=1/b(E_p)$, could not be clearly identified for energies $E_p$ close to $E_{p,+}$ (data not shown, but see Fig.~\ref{fig:conformational}(b) for an example of such a $b(E_p)$ curve).
	{Here it is worth mentioning that we also include in Fig.~\ref{fig:phasediagram}(a) the results extracted from Ref.~\cite{campa2016jstat} (black crosses), which, despite of being obtained from a different approach, are in accordance with our results.
	In that reference, it was also estimated that the canonical critical point for $\alpha=2$ is at $\eta_{\text{cp}}^{\,}=2$, which means that, for $\eta<\eta_{\text{cp}}^{\,}$ ({\it i.e.}, $\rho > \rho_{\text{cp}}^{\,}$), there is no first-order phase transition in the canonical ensemble and the temperatures $T^*$, $T_-$ and $T_+$ converge to the same value at $\rho_{\text{cp}}^{\,} \approx 0.12$, as it is also suggested by our results in Fig.~\ref{fig:phasediagram}(a).

\begin{figure*}[!b]
\centering
\includegraphics[width=0.95\textwidth]{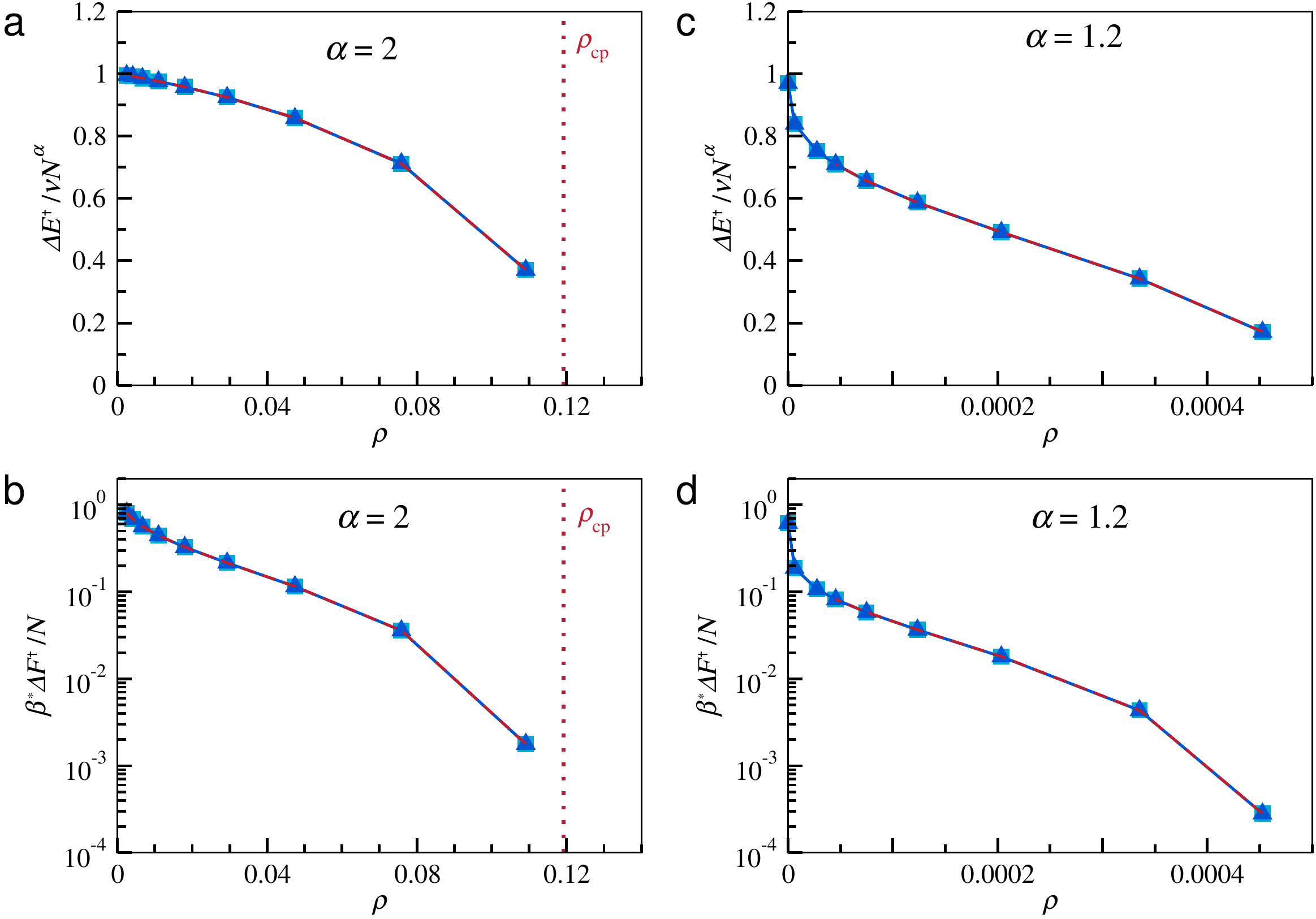}
\caption{Top panels (a) and (c) show the rescaled latent heats, $\Delta E^{\dagger}/N$, while bottom panels (b) and (d) display the rescaled free-energy barriers, $\beta^{*}\Delta F^{\dagger}/N$, as functions of the concentration $\rho$, for $\alpha=2$ (left panels) and $\alpha=1.2$ (right panels).
Filled symbols indicate the results obtained via the usual microcanonical analysis:
the squares correspond to data obtained for $\nu=5$ and $N=5\times 10^3$, while the triangles correspond to data obtained for $\nu=9$ and $N=10^4$. 
Short-dashed (red) lines denote the results obtained using the approach based on the conformational microcanonical entropy.
Vertical dotted lines in (a) indicate the critical point $\rho_{\text{cp}}^{\,}$ 
extracted from Ref.~\cite{campa2016jstat}.
All results are displayed assuming $k_B=1$.
}
\label{fig:latheat-deltaF}
\end{figure*}

\subsection{Latent heat and free-energy barriers}
\label{barrier}

	Now, in order to further characterize the first-order phase transitions observed in our aggregation model, we include in Fig.~\ref{fig:latheat-deltaF} the rescaled latent heats, $\Delta E^\dagger/\nu N^\alpha$, and the free-energy barriers per particle, $\beta^*\Delta F^\dagger/N$, as functions of $\rho$ for both $\alpha=2$ and $\alpha=1.2$.
	We consider the same set of parameters $N$ and $\nu$ used to produce Fig.~\ref{fig:phasediagram}, with squares denoting results for $\nu=5$ and $N=5\times 10^3$, and triangles the results for $\nu=9$ and $N=5\times 10^4$.
	Accordingly, the latent heat and the free-energy barrier display the same scaling behaviour observed, respectively, for the energy $E$ and the free-energy profile $\beta^{*}\Delta F(E)$ in Figs.~\ref{fig:alpha2}(d) and~\ref{fig:alpha12}(d).

	In addition, the results in Fig.~\ref{fig:latheat-deltaF} suggest that, for both $\alpha=2$ and $\alpha=1.2$, the latent heat and free-energy barrier decrease as the concentration $\rho$ increases.
	As expected from the canonical critical point for $\alpha=2$ evaluated 
in Ref.~\cite{campa2016jstat} (and mentioned in the previous section), both should be zero 
at $\eta_{\text{cp}}^{\,}=2$ ({\it i.e.}, $\rho_{\text{cp}}^{\,} \approx 0.12$).
	Indeed, Fig.~\ref{fig:latheat-deltaF}(b) indicate that the free-energy barrier decreases
very rapidly from $\sim N$, at low concentrations, to $\sim 0$, as $\rho$ get closer to the critical value.
	The same qualitative behaviour is observed for $\alpha=1.2$ in Fig.~\ref{fig:latheat-deltaF}(d).
	Figures~\ref{fig:latheat-deltaF}(a) and~\ref{fig:latheat-deltaF}(c) show that the latent heats also decrease with concentration, but in a smoother way.
	As indicated by the data displayed as short-dashed (red) lines in Fig.~\ref{fig:latheat-deltaF}, the results determined from the usual microcanonical analysis are in good agreement with
the results obtained via the conformational ensemble analysis discussed in Sec.~\ref{conformational}.


\section{Conclusions}

	In this work we have considered a generalized model inspired on the Thirring's model which allowed us to identify shape-free thermostatistics properties of the aggregation phase transitions.
 	By considering a microcanonical analysis method, we have characterized the equilibrium thermostatistics for both the usual and the generalized version of the model, showing that they 
present first-order phase transitions for sufficiently high values of the control parameter
$\eta$, {\it i.e.}, low concentrations $\rho$.
	From the microcanonical characterization it was possible to obtain not only the transition 
 temperatures $T^{*}$, and the spinodal ($T_{-}$) and solubility lines ($T_{+}$), but also the latent heats $\Delta E^{\dagger}$ and the free-energy barriers $\beta^{*} \Delta F^{\dagger}$, as a function of the concentration $\rho$.
	We also explored how the changes in the other parameters of the system, namely, the total number of particles $N$ and the magnitude of interaction energy between aggregate particles $\nu$, affect the thermostatistics of the system. 
	Despite of the non-additivity of the model, our results indicate that the thermodynamic quantities such as temperature, entropy, and energy, obey simple but non-trivial relationships with the parameters $N$ and $\nu$.
	We note that further analyses obtained for $\alpha=1.4$, $1.6$, and $1.8$ (data not shown) suggest that 
such scaling relations may be also valid for other values within the range $1 < \alpha \leq 2$.
	In addition, by considering the conformational microcanonical ensemble, we show that
one can obtain the same phase diagrams, latent heats, and free-energy barriers, as the description extracted 
from the usual microcanonical approach which takes into account the contribution of kinetic energies,
even though the obtained caloric curves and free-energy profiles are different.

	Finally, it is worth mentioning that our semi-analytical results might have an important 
influence not only on the microcanonical analyses of first-order phase transitions
in finite-sized systems with disordered aggregates obtained from Monte Carlo simulations (see, {\it e.g.}, Refs.~\cite{zierenberg2014jcp,mueller2015physproc,zierenberg2018jphysconfser,janke2017natcommun,schierz2016pre,janke2017jphysconfser,schierz2015jcp,trugilho2020jphysconfser}), but also  on the recently developed kinetic approaches
which use the microcanonical entropies to estimate temperature-dependent rate constants~\cite{rizzi2020jstat,trugilho2021arxiv}.

\vspace{0.5cm}

\noindent
{\small \bf Acknowledgements}
{\small The authors acknowledge the financial support from the Brazilian agencies CAPES (code 001), FAPEMIG 
(Process\,APQ-02783-18), and CNPq (Grants N$^{\b{o}}$ 306302/2018-7 and N$^{\b{o}}$ 426570/2018-9).}

\vspace{-0.2cm}







%

\end{document}